\documentstyle[twocolumn,graphicx,floats,prl,aps]{revtex}
\tighten

\draft
\begin{document}
\wideabs{
\title{Interaction of an electron gas with photoexcited
electron-hole pairs in modulation-doped GaAs and CdTe
quantum wells}

\author{H. A. Nickel,$^{\dag}$\footnote{\small Corresponding author: 239 Fronczak
Hall, Buffalo, NY 14260, USA, email: nickel@buffalo.edu} 
        T. Yeo,$^{\dag}$ C. J. Meining,$^{\dag}$
	D. R. Yakovlev,$^{\ddag}$\cite{DRY}
        M. Furis,$^{\dag}$ A. B. Dzyubenko,$^{\dag}$\cite{ABD} \\
	B. D. McCombe,$^{\dag}$ and A. Petrou$^{\dag}$}
\address{$^{\dag}$Department of Physics, University at Buffalo, SUNY,
Buffalo, NY 14260, USA}
\address{$^{\ddag}$Physikalisches Institut der Universit\"at W\"urzburg,
D-97074 W\"urzburg, Germany}
\date{\today}
\maketitle
\begin{abstract}
The nature of the correlated electron gas and its response
to photo-injected electron-hole pairs in nominally undoped and
modulation-doped multiple quantum-well (MQW) structures was studied
by experiment and theory, revealing a new type of optically-active
excitation, magnetoplasmons bound to a mobile valence hole.  These
excitations are blue-shifted from the corresponding transition of the
isolated charged magnetoexciton $X^-$.  The observed blue-shift of $X^-$
is larger than that of two-electron negative donor $D^-$, in agreement
with theoretical predictions.
\end{abstract}

} 

Since the initial observation of the negatively charged exciton $X^-$
in  CdTe/CdZnTe QWs \cite{kheng93a} there has been considerable interest
in charged electron-hole ($e$-$h$) complexes in quasi-two-dimensional
semiconductor systems.  $X^-$, the semiconductor analog to the
negatively charged hydrogen ion $H^-$ in atomic physics, differs from its
superficially similar relative, the negatively charged donor ion $D^-$
\cite{jiang97a} in some very important respect: the positive charge in
$X^-$, the hole, is free to move around. A symmetry associated with the
resulting center-of-mass motion leads to a new electric-dipole selection
rule \cite{dzyubenko00b}, which prohibits, in particular, certain
bound-to-bound $X^-$ transitions \cite{dzyubenko00a} that dominate the
absorption spectra of $D^-$.  The evolution of the spectra in 2D $e$-$h$
systems with electron density $n_e$ from isolated neutral excitons, via
isolated $X^-$, to a plasma consisting of many electrons and a few holes,
and the effects of magnetic fields on this evolution have been the subject
of many studies.  Magneto-photoluminescence (PL) spectra of such systems
show bandgap-renormalized Landau-level (LL)-to-LL transitions at high
filling factors (FF) $\nu=2\pi l_B^2 n_e$, $l_B=(\hbar c/eB)^{1/2}$ and
a discontinuity in the slope of the PL energy vs. field for the $N=0$
LL transition at $\nu=2$. For $\nu<2$ the dominant PL is depressed
below the extrapolated $N=0$ LL transition and has an ``exciton-like''
field dependence (see, e.g., \cite{gekhtman96a,rashba00} and references
therein).  Internal transitions of excitons (IETs) can yield important
insight into the ground and excited states of excitons \cite{salib96a}
and thus offer another method of probing the excitonic state in the
dilute situation and its evolution with $n_e$ and magnetic field. In
this study we have used the internal transitions of negatively charged
excitons $X^-$ \cite{dzyubenko00a} and the effects of many-electrons on
these spectroscopic signatures to probe the many-body state of the system.

Three GaAs/Al$_{0.3}$Ga$_{0.7}$As MQW samples (samples 1, 2, and 3) and
one CdTe/Cd$_{0.7}$Mg$_{0.3}$Te MQW sample (sample 4) were studied.
The structures of these samples are [(well/barrier--thickness in
\AA)$\,\times\,$repetitions]: sample 1 -- (200/600)$\,\times\,$20;
sample 2 -- (240/480)$\,\times\,$20, modulation-doped at
8.0$\times$10$^{10}$cm$^{-2}$; sample 3 -- (240/240)$\,\times\,$10,
modulation-doped at 3.2$\times$10$^{11}$cm$^{-2}$; and sample 4 --
(80/330)$\,\times\,$10, modulation-doped at 3.5$\times$10$^{11}$cm$^{-2}$.
The dopant in the GaAs (CdTe) samples was silicon (iodine).  These
structures were studied by photoluminescence (PL) and optically detected
resonance (ODR) spectroscopy \cite{dzyubenko00a,salib96a} at low (4.2K)
temperatures in magnetic fields up to 15~Tesla.  The range of doping
densities chosen for this experiment puts the magnetic field corresponding
to FF $\nu=2$ well within the range of the superconducting magnet used
in this study. The observed resonances were studied at several discrete
far-infrared (FIR) laser photon energies ranging from 2.87 meV to
17.6 meV.

%
%
\begin{figure}[tb]
\begin{center}
\begin{minipage}{0.99\linewidth}
\begin{center}
\includegraphics[clip=,angle=-90,width=0.98\textwidth]{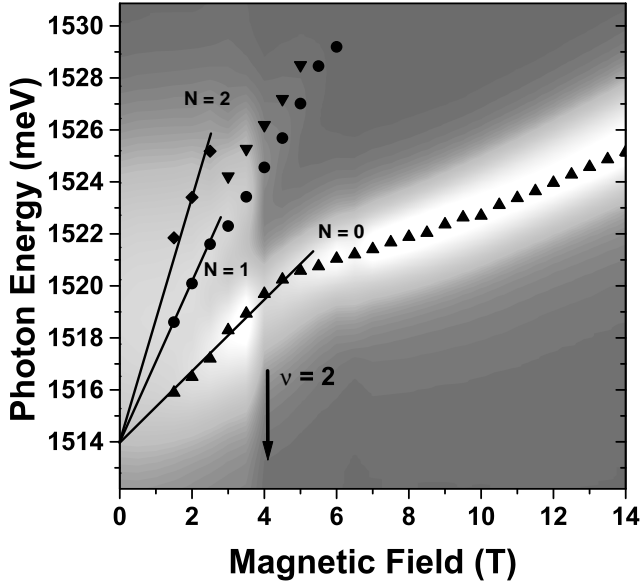}
\end{center}
\vspace*{-1ex}
\caption{Magneto-photoluminescence spectrum of sample 3. Datapoints
were obtained by peak-fitting individual spectra, solid lines are
guides to the eye.}  \label{fig1}
\end{minipage}
\end{center}
\end{figure}
Figure~1 shows magneto-photoluminescence (MPL) data for sample 3.
Overlaying the contour plot  are datapoints indicating peak positins that
were obtained by peak-fitting  the individual MPL spectra. The solid
lines indicating LL-to-LL recombination are guides to the eye. After
peak-fitting the MPL data, an ODR spectrum was obtained by recording
the intensity changes (induced by the absorption of FIR radiation) of
the main PL feature (indicated by the upright triangles) as a function
of magnetic field.  Figure~2a shows ODR spectra for sample~2 for several
FIR laser photon energies as indicated in the graph. The magnetic field
corresponding to FF $\nu = 2$, 1.66~Tesla, is indicated by the arrow at
the bottom axis. The sharp feature present in all traces and marked by
the upward pointing arrowhead is identified as the electron cyclotron
resonance ($e$-CR) transition with energy $\hbar\omega_{\rm ce}$. Note
that no feature is observed in any scan at a higher magnetic field than
that of $e$-CR, in agreement with the selection rule \cite{dzyubenko00b},
prohibiting bound-to-bound IETs of $X^-$. Two other features are
observed in the ODR spectra recorded for high FIR photon energies. These
features, which are blue-shifted from those observed in the undoped
GaAs sample, are attributed to bands of ionizing singlet- (circles) and
triplet-like (crosses) internal transitions of $X^-$ \cite{dzyubenko00a}.
The resonant fields for both singlet- and triplet-like $X^-$ transitions
occur {\em above\/} the field corresponding to $\nu =2$.  At lower FIR
photon energies, the triplet feature remains strong, but the singlet
feature weakens and is not observable at 6.73\,meV (the predicted field
position corresponds to $\nu > 2$).  For the heavily doped sample 3
these features become generally weaker and are further blue-shifted.
In the modulation-doped CdTe-based MQW (Fig.~2b), only a triplet-like
band below $e$-CR was observed for FF $\nu < 2$.  The field position
of this feature is also blue-shifted with respect to the position in a
nominally undoped sample.

%
%
\begin{figure}[tb]
\begin{center}
\begin{minipage}{0.99\linewidth}
\begin{center}
\includegraphics[clip=,angle=-90,width=0.98\textwidth]{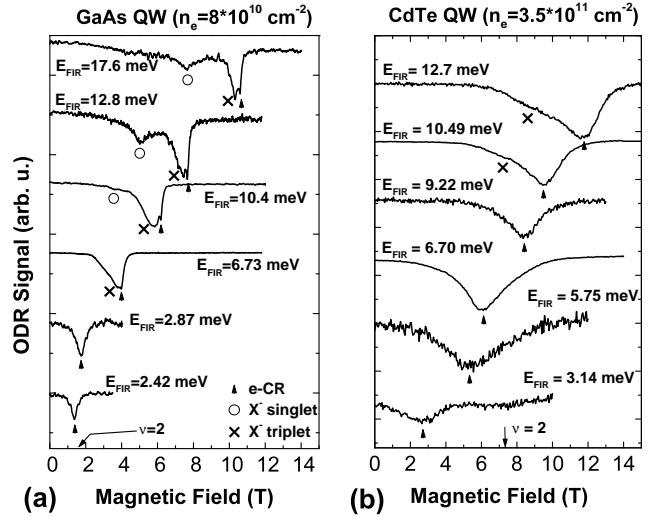}
\end{center}
\vspace*{-1ex}
\caption{ODR spectra of sample 2 (GaAs) and 4 (CdTe).} \label{fig2}
\end{minipage}
\end{center}
\end{figure}

Theoretically, we consider intraband excitations from the ground
state of a 2D system ``the hole embedded in a sea of electrons''
in strong magnetic fields with integer fillings of electron LL's
$\nu=1, 2$. In the presence of a compensating background, intraband
collective excitations from the ground state can be described to be
resonances involving a correlated $e$-$h$ final state consisting of
the conduction band electron in the empty $N=1$ LL, $e_{\rm 1cb}$,
a conduction band hole in an otherwise filled $N=0$ LL, $h_{\rm
0cb}$, and a mobile valence band hole in the $N=0$ LL, $h_{\rm 0vb}$.
Therefore, the final state can be considered to be a positively charged
$2h$-$e$ complex, which closely resembles the positively charged trion
$X^+$. The peculiarity of the present situation is the strong exchange
interaction between the particles $e_{\rm 1cb}$ and $h_{\rm0cb}$.
In the absence of the valence band hole, such a collective excitation
is a 2D magnetoplasmon \cite{bychkov81,kallin84} Magnetoplasmons,
like charge-neutral excitations, can be characterized by the conserved
center-of-mass momentum ${\bf K}=(K_x,K_y)$. When $|{\bf K}|$ ranges from
zero to infinity, the magnetoplasmon dispersion $E({\bf K})$ forms a band
of finite width $E_0/2$, where $E_0=\sqrt{\pi/2}\, e^2/\epsilon l_B$.
An eigenstate of the magnetoplasmon with a fixed {\bf K} is characterized
by the finite relative motion (a bound $e_{\rm 1cb}$-$h_{\rm 0cb}$ pair)
and by the extended motion of the center-of-mass.  In the presence of
the valence band hole, the classification of states changes drastically:
the three-particle state $X^+$ is charged and cannot be characterized
by the conserved center-of-mass momentum ${\bf K}$.  Instead, charged
complexes in ${\bf B}$ can be characterized \cite{dzyubenko00b} by two
exact quantum numbers ---  the total angular momentum projection $M_z$
and the oscillator quantum number $k$. The latter physically determines
the center-of-rotation of the charged complex as a whole in the plane
perpendicular to ${\bf B}$; correspondingly, there is Landau degeneracy
in $k$.  The relative motions of three particles can be either finite
(on average, all particles are at finite distances from each other)
or infinite. In the latter case the states belong to the three-particle
continuum, which is formed by two partly overlapping bands:  one band is
formed by the magnetoplasmon (the bound $e_{\rm 1cb}$-$h_{\rm 0cb}$ pair)
plus a valence band hole $h_{\rm 0vb}$ in a scattering state. The second
band is formed by the states of an interband magnetoexciton $X_{10}$
(a bound $e_{\rm 1cb}$-$h_{\rm 0vb}$ pair) plus a conduction-band
hole $h_{\rm 0cb}$ in the scattering state. When all relative motions
are finite, a truly bound $X^+$-like state is formed; such states lie
outside the continuum. There also exists a third possibility: quasi-bound
states --- the three-particle $2h$-$e$ resonances --- may exist in the
continuum. Depending on the width of the resonances and on the coupling
to the continuum, such states may resemble bound $X^+$ states and may
exhibit sharp peaks or Fano-resonances in optical measurements.

We performed calculations of the eigenspectra of the three-particle
$2h$-$e$ excitations and matrix elements of intraband optical transitions
$h_{\rm 0vb} + {\sl photon} \rightarrow 2h$-$e$ following the method
that has been described elsewhere \cite{dzyubenko00b}.  In this paper,
we describe the spectra of transitions that have energies larger than
the $e$-CR energy $\hbar\omega_{\rm ce}$.  For electron FF $\nu=1$,
there exists one rather sharp, optically active $2h$-$e$ resonance
that lies within the continuum. For FF $\nu=2$, due to the enhanced
exchange-correlation effects, this state shifts upward in energy,
moves out of the continuum, and becomes a truly bound $2h$-$e$ state.
Simultaneously, it loses oscillator strength.  Physically, for both
$\nu=1$ and $\nu=2$, the optically active state describes a magnetoplasmon
bound to the mobile valence band hole $h_{\rm 0vb}$. The corresponding
energies of collective excitations are larger than the energy of the
internal triplet $X^-$ transition (see inset to Fig.~3). The latter
transition corresponds to the vanishing FF $\nu=0$. Therefore, the
$X^-$ transitions are blue-shifted in the presence of excess electrons.
The calculated blue shifts of the $X^-$ triplet at $\nu=1$ and $\nu=2$
are 0.28$E_0$ and 0.49$E_0$, correspondingly.

%
%
\begin{figure}[tb]
\begin{center}
\begin{minipage}{0.99\linewidth}
\begin{center}
\includegraphics[clip=,angle=-90,width=0.98\textwidth]{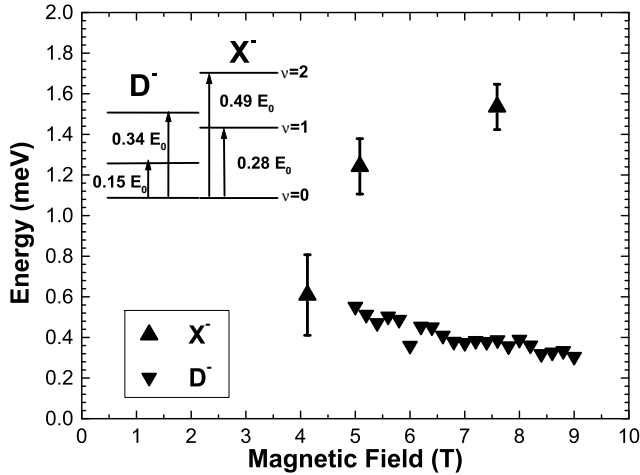}
\end{center}
\vspace*{-1ex}
\caption{
Summary plot showing the blue shift of
$X^-$ and $D^-$.  $E_0=\protect\sqrt{\pi/2}\, e^2/\epsilon l_B$
is the characteristic energy of Coulomb interactions in
strong magnetic fields.}  \label{fig3}
\end{minipage}
\end{center}
\end{figure}
This behavior resembles the blue shift of the two-electron negative donor
center $D^-$ in the presence of excess electrons \cite{cheng93}, which
is understood in terms of magnetoplasmons localized on fixed donor ions
$D^+$ \cite{dzyubenko93a,hawrylak94a}. The qualitative difference for a
mobile valence hole is the presence of an additional dynamical symmetry,
magnetic translations, and the corresponding exact optical selection
rule \cite{dzyubenko00b,dzyubenko00a}.  Quantitatively, the blue shifts
of the $X^-$ and $D^-$ are comparable. In agreement with experiment,
the predicted blue shift of the $X^-$ is larger than that of the $D^-$
(Fig.~3).  This can be explained by the diminished negative contribution
of the Coulomb $e$-$h$ attraction to the energy of the final $2h$-$e$
state containing a mobile valence band hole.

In conclusion, we have theoretically shown that in a photo-excited,
quasi-2D system, in the presence of an electron gas, there exists a
new type of optically active excitation, a magnetoplasmon bound to a
mobile valence hole, and we have obtained experimental evidence for
this transition in ODR experiments. This excitation, depending on the
electron filling factor $\nu$, lies either within or above the band of
free neutral magnetoplasmons and is blue shifted from the corresponding
transition of the isolated charged magnetoexciton $X^-$.  Theoretical
results provide a qualitative explanation of the experimental findings,
the blue shift of internal $X^-$ transitions for fillings factors $\nu <
2$, in terms of a collective many-body excitation of the system.

We thank W.~Schaff, T.~Wojtowicz, and G.~Karczewski for the excellent
MBE growth of the GaAs and CdTe samples used in this study.  This work
was supported in part by the NSF grant DMR 9722625 and by the COBASE
grant R175.



\vspace*{-3ex}
\begin{references}
\vspace*{-6ex}
\bibitem[**]{DRY}On leave from A.~F.~Ioffe Physico-Technical Institute,
		 RAS, St. Petersburg 194017, Russia

\bibitem[***]{ABD}On leave from General Physics Institute, RAS,
                Moscow 117942, Russia.

\bibitem{kheng93a}
             K.~Kheng, R.~T.~Cox, Y.~Merle d'Aubigne,
             F.~Bassani, K.~Saminadayar, and S.~Tatarenko,
             Phys. Rev. Lett. {\bf 71}, 1752 (1993).

\bibitem{jiang97a}
  Z.~X.~Jiang, B.~D.~McCombe, Jia-Lin~Zhu, and W.~Schaff,
  Phys. Rev. B {\bf 56}, R1692 (1997).

\bibitem{dzyubenko00b}
A.~B.~Dzyubenko and A.~Yu.~Sivachenko, Phys. Rev. Lett. {\bf 84}, 4429 (2000).

\bibitem{dzyubenko00a}
             A.~B.~Dzyubenko, A.~Yu.~Sivachenko, H.~A.~Nickel,
             T.~M.~Yeo, G.~Kioseoglou, B.~D.~McCombe, and A.~Petrou,
             Physica E {\bf 6}, 156 (2000).



\bibitem{gekhtman96a}
D.~Gekhtman, E.~Cohen, Arza Ron, and L.~N.~Pfeiffer,
Phys. Rev. B {\bf 54}, 10320 (1996).

\bibitem{rashba00}
E.~I.~Rashba and M.~D.~Sturge, Phys. Rev. B {\bf 63}, 045305 (2001).

\bibitem{salib96a}
                     M.~Salib, H.~A.~Nickel, G.~S.~Herold, A.~Petrou,
 		     B.~D.~McCombe, R.~Chen, K.~K.~Bajaj, and W.~Schaff,
                     Phys. Rev. Lett. {\bf 77}, 1135 (1996);
               J.~\v{C}erne, J.~Kono, M.~S.~Sherwin, M.~Sundaram,
	       A.~C.~Gossard, and G.~E.~W.~Bauer,
               Phys. Rev. Lett. {\bf 77}, 1131 (1996).

\bibitem{bychkov81}
Yu.~A.~Bychkov, S.~V.~Iordanskii, and G.~M.~Eliashberg,
        Pis'ma Zh. Eksp. Teor. Fiz. {\bf 33}, 152 (1981)
	[JETP Lett.  {\bf 33},  143 (1981)].

\bibitem{kallin84}
C.~Kallin and B.~I.~Halperin, Phys. Rev. B {\bf 30},  5655 (1984).

\bibitem{cheng93}
J.-P.~Cheng, Wang Y.~J.~Wang , B.~D.~McCombe, and W.~Schaff,
Phys. Rev. Lett. 70, 489 (1993).


\bibitem{dzyubenko93a}
A.~B.~Dzyubenko and Yu.~E.~Lozovik,
        Zh. Eksp. Teor. Fiz. {\bf 104}, 3416 (1993)
	[JETP {\bf 77}, 617 (1993)].

\bibitem{hawrylak94a}
P.~Hawrylak, Phys. Rev. Lett. {\bf 72}, 2943 (1994).
\end{references}
\end{document}